\documentclass[12pt]{article}

\pdfoutput=1

\usepackage[top=80pt,bottom=85pt,left=85pt,right=85pt]{geometry}
\usepackage{amsmath,amssymb,graphicx,float,color,tikz,subfigure,sectsty,cite}
\usepackage[debug,pageanchor=false]{hyperref}
\definecolor{link}{rgb}{.8,.15,.1}
\hypersetup{colorlinks=true,linkcolor=link,citecolor=link,urlcolor=link,linktocpage}

\newcommand{\de}{\partial} 
\def\dd{\mathrm{d}}

\def\im{\mathrm{Im}}
\def\vf{\varphi}

\def\a {\hat a}
\def\b {\hat b}
\def\t {\hat c}

\makeatletter
\@addtoreset{equation}{section}
\makeatother

\begin{document}

	\begin{titlepage}
	
	\begin{flushright}
        {\tt UUITP-29/18}\\
        \end{flushright}

	\begin{center}

	\vskip .5in 
	\noindent

	{\Large \bf{AdS$_3$ solutions with exceptional supersymmetry}}

	\bigskip\medskip

	Giuseppe Dibitetto,$^1$ Gabriele Lo Monaco,$^2$ Achilleas Passias,$^1$ Nicol\`o Petri,$^3$ Alessandro Tomasiello$^2$\\

	\bigskip\medskip
	{\small 

	$^1$ Department of Physics and Astronomy, Uppsala University, \\ Box 516, SE-751 20 Uppsala, Sweden
	\\	
	\vspace{.3cm}
	$^2$ Dipartimento di Fisica, Universit\`a di Milano--Bicocca, \\ Piazza della Scienza 3, I-20126 Milano, Italy \\ and \\ INFN, sezione di Milano--Bicocca
	\\	
	\vspace{.3cm}
	$^3$ Department of Mathematics, Bo\u{g}azi\c{c}i University, 34342, Bebek, Istanbul, Turkey
	}

	\vskip .5cm 
	{\small \tt giuseppe.dibitetto@physics.uu.se, g.lomonaco1@campus.unimib.it, achilleas.passias@physics.uu.se, nicolo.petri@boun.edu.tr, alessandro.tomasiello@unimib.it}
	\vskip .9cm 
	     	{\bf Abstract }
	\vskip .1in
	\end{center}

	\noindent
	
Among the possible superalgebras that contain the AdS$_3$ isometries, two interesting possibilities are the exceptional $F(4)$ and $G(3)$. Their R-symmetry is respectively SO(7) and $G_2$, and the amount of supersymmetry ${\cal N}=8$ and ${\cal N}=7$. We find that there exist two (locally) unique solutions in type IIA supergravity that realize these superalgebras, and we provide their analytic expressions. In both cases, the internal space is obtained by a round six-sphere fibred over an interval, with an O8-plane at one end. The R-symmetry is the symmetry group of the sphere; in the $G(3)$ case, it is broken to $G_2$ by fluxes. We also find several numerical ${\cal N}=1$ solutions with $G_2$ flavor symmetry, with various localized sources, including O2-planes and O8-planes.

	\noindent

	\vfill
	\eject

	\end{titlepage}

	\tableofcontents

\section{Introduction} 
\label{sec:intro}

Among conformal field theories (CFTs), two-dimensional ones have a privileged role, as the conformal algebra in two dimensions is infinite-dimensional. From a holographic point of view, however, they are in fact harder to classify: AdS$_3$ solutions in string theory have an internal space $M_7$ of dimension seven, which is large enough to open a plethora of possibilities (see for example \cite{Boonstra-Skenderis, argurio-giveon-shomer, Kim, Gauntlett-Kim-Waldram-1, Gauntlett-Conamhna-Waldram-2, Donos, Couzens-Martelli-Schafer-Nameki1, Eberhardt}). This is to be contrasted with CFTs in higher dimensions, where the internal space has fewer dimensions, its geometry is tightly constrained, and a classification easier.

Whenever faced with a problem which offers too many possibilities, one can achieve progress by imposing additional symmetries. Imposing supersymmetry is a natural choice. For AdS$_3$, however, the minimal amount corresponds to only two supercharges, which are unlikely to constrain much the dynamics. This suggests looking at extended supersymmetry, i.e.~${\cal N}>1$. In this situation, the conformal algebra $\mathfrak{so}(2,1)$ is complemented not only by the supercharges, but also by additional bosonic generators (the so-called R-symmetry). Together, all these generators span an interesting superalgebra.
   
Looking at the list of possible superalgebras that contain a $\mathfrak{so}(2,1)$ summand, one finds two particularly intriguing possibilities: the exceptional superalgebras $F(4)$ and $G(3)$.\footnote{The recent work \cite{beck-gran-gutowski-papadopoulos} found that these are allowed superalgebras for an AdS$_3$ solution; see Table 2 there. In older work, superconformal algebras in two dimensions associated with $F(4)$ and $G(3)$ were found and studied \cite{fradkin-linetsky,fradkin-linetsky2,gunaydin-ketov}, confirming this possibility from the dual CFT point of view.} These correspond to ${\cal N}=8$ and $7$ respectively; the R-symmetry algebras are $\mathfrak{so}(7)$ and $\mathfrak{g}_2$. Exceptional algebras appear in several contexts in string theory and mathematical physics, and they are usually a beacon for interesting phenomena. 

In this paper, we will look for AdS$_3$ solutions realizing these two exceptional superalgebras. Their R-symmetry should be realized as a symmetry of the internal space; for SO(7), the orbits of its action can either be $\mathrm{SO}(7)/G_2\cong \mathbb{RP}^7$ or $\mathrm{SO}(7)/\mathrm{SO}(6)\cong S^6$. The former would mean that the entire internal space is a coset, and unfortunately does not work; thus we focus on the latter. $S^6$ does not fill the entire internal space, but rather has codimension one; we thus have to fiber it over an interval. In this situation $M_7$ is said to have cohomogeneity one. Similarly, for a $G_2$ R-symmetry, the only possible orbits are $G_2/\mathrm{SU}(3)$. This is again a round $S^6$, which again we have to fiber over an interval. This time, however, the coset realization points to the existence of forms which break the SO(7) isometries of the round $S^6$ to $G_2$. We will use these forms as internal fluxes.

For both $F(4)$ and $G(3)$ we will find analytic solutions. Their maximal extension without D8-branes is noncompact, leading to infinite central charge. Similar to AdS$_7$ solutions \cite{afrt}, we explore the possibility of cutting and gluing them along D8-branes. The central charge becomes then finite, and intriguingly it displays the $k^{1/3} N^{5/3}$ behavior  \cite{ajtz,jafferis-klebanov-pufu-safdi,guarino-jafferis-varela} of the free energy for AdS$_4$ solutions with Romans mass.

Having found solutions with exceptional superalgebras, we also looked for solutions whose \emph{flavor} symmetry is exceptional. In particular, since we learned how to realize $G_2$ as the R-symmetry, we decided to use a similar strategy to realize $G_2$ as a flavor symmetry. 
In this setup, the $\mathbf{8}$ of SO(7) is branched into the $\mathbf{1}\oplus\mathbf{7}$ of $G_2$, and thus, besides the possibility of preserving $\mathcal{N}=7$ or $8$ (both cases yielding trivial flavor symmetry), we can preserve $\mathcal{N}=1$, in which case $G_2$ flavor symmetry is realized.
As expected, this problem is far less rigid than the realization of exceptional superalgebras; still it is constrained enough to make progress. We obtained several numerical solutions with fully localized O8- and O2-planes. 

It is interesting to notice that type II AdS$_d$ solutions with an internal $S^{9-d}$ fibred over an interval now exist for several values of $d$: for $d=6$ \cite{brandhuber-oz}, $d=7$ \cite{afrt}, $d=2$ \cite{Dibitetto:2018gbk} and $d=3$ in this paper, perhaps suggesting a pattern.

We will start in section \ref{sec:ex} with a quick review of the two exceptional superalgebras we are interested in, and outlining a strategy for realizing them in supergravity. In section \ref{sec:setup} we will translate that into a concrete Ansatz for the metric and fluxes; we will also review the relevant aspects of the pure spinor formalism we will use to impose supersymmetry. We will then proceed to find the explicit analytic solutions. For $F(4)$, we will actually first find a solution in section \ref{sub:dbhat} using a near-horizon limit from a certain brane system, and then recover it in section \ref{sub:f4susy} using pure spinors, thereby showing its local uniqueness. For $G(3)$, we will use directly the pure spinor method in section \ref{sec:g3}. Finally, in section \ref{sec:g2} we will find numerical solutions with ${\cal N}=1$ supersymmetry and $G_2$ flavor symmetry.


\section{Exceptional superalgebras} 
\label{sec:ex}

We will start with some introductory considerations on how to realize the superalgebras we are interested in. 
The exceptional superalgebra $F(4)$ has a bosonic subalgebra
\begin{equation}\label{eq:f4bos}
	\mathfrak{sl}(2) \oplus \mathfrak{so}(7)\,,
\end{equation}
and sixteen fermionic generators in the $(\mathbf{2}, \mathbf{8})$ representation. It has appeared in the context of six-dimensional supergravity \cite{romans-F4}, with a different real form whose bosonic subalgebra is $\mathfrak{so}(5,2)\oplus \mathfrak{su}(2)$. In that case, the summand $\mathfrak{so}(5,2)$ is interpreted as the isometry algebra of AdS$_6$, while $\mathfrak{su}(2)$ is an R-symmetry. In the present context, the opposite will be the case: the summand $\mathfrak{sl}(2)$ will be taken to be the $\mathfrak{sl}(2)_\mathrm{L}$ of the $\mathfrak{so}(2,2)\cong\mathfrak{sl}(2)_\mathrm{L}\oplus \mathfrak{sl}(2)_\mathrm{R}$ isometry algebra of AdS$_3$, while $\mathfrak{so}(7)$ will be the R-symmetry algebra. 

To realize this superalgebra for an AdS$_3$ solution, we then have to look for solutions with an $\mathfrak{so}(7)$ symmetry algebra. The corresponding group will be $G=$ SO(7) or Spin(7). On its orbits (the sets obtained by acting on a point by $G$), the action of $G$ is by definition transitive; since our internal manifold $M_7$ has dimension seven, the orbits need to have dimension $\le 7$. Calling $H$ the little group of any point, we see that each orbit is of the form $G/H$ for some $H$. The possible maximal subgroups of $G=\mathrm{SO}(7)$ are, up to discrete quotients, $\mathrm{SO}(6)$, $\mathrm{SU}(2)^3$, $\mathrm{SO}(5)\times \mathrm{U}(1)$, and $G_2$. Among these, only for $H=\mathrm{SO}(6)$ and $G_2$ we have a candidate orbit $G/H$ of dimension $\le 7$. (We need not consider non-maximal subgroups: they will have even smaller dimension, and $G/H$ will have even higher dimension.) So we have two candidate orbits, up to discrete quotients:
\begin{equation}\label{eq:f4orbits}
	\mathrm{Spin}(7)/G_2\cong S^7 \, ,\qquad \mathrm{SO}(7)/\mathrm{SO}(6)\cong S^6\,.
\end{equation}
(In the first case we have taken $G=\mathrm{Spin}(7)$ rather than SO(7) to obtain a simply connected orbit, for simplicity.) The metrics compatible with these cosets are the round ones.

The simplest possibility in (\ref{eq:f4orbits}) might perhaps seem taking $M_7=S^7$; that would mean that the orbit coincides with the whole space. Unfortunately, we will show rather quickly in section \ref{sub:susy} below that this does not work. 

This leaves us with the possibility of taking the orbits equal to $S^6$. This is the one we will pursue in this paper; as we will see, it does lead to solutions. 

Let us now consider the fermionic generators, i.e.~the supercharges. These will be realized by the IIA supersymmetry transformations with supersymmetry parameters
\begin{equation}\label{eq:eps}
	\epsilon_1 = \sum_{I=1}^{\cal N}\zeta^I \otimes {1 \choose - i} \otimes \eta^I_1
	\ ,\qquad
	\epsilon_2 = \sum_{I=1}^{\cal N}\zeta^I \otimes {1 \choose + i} \otimes \eta^I_2\, ,                                           
\end{equation}
where $\eta_{1,2}^I$ are seven-dimensional Majorana spinors; as usual for AdS solutions, $\zeta^I$ are three-dimensional Majorana Killing spinors: 
\begin{equation}\label{eq:Ksp}
\nabla_\mu \zeta = \frac{1}{2} \mu \gamma_\mu \zeta\,.
\end{equation}
Since we want the supercharges to transform in the $\mathbf{8}$ of $\mathfrak{so}(7)$, we need to take ${\cal N}=8$; in other words, $I$ becomes a spinorial SO(7) index. Moreover, there are two independent solutions to (\ref{eq:Ksp}), which transform in the $\mathbf{2}$ of $\mathfrak{sl}(2)_\mathrm{L}$ and are singlets under $\mathfrak{sl}(2)_\mathrm{R}$. (In (\ref{eq:eps}), each $\zeta^I$ can be taken to be any linear combination of these two.) Given these transformations laws, the supercharges obtained from (\ref{eq:eps}) complete $\mathfrak{sl}(2)_\mathrm{L}\oplus \mathfrak{so}(7)$ to the superalgebra $F(4)$, while $\mathfrak{sl}(2)_\mathrm{R}$ is left undisturbed.\footnote{On AdS$_3$ there are also solutions to the equation obtained from (\ref{eq:Ksp}) by reversing the sign of $\mu$; these would transform in the $\mathbf{2}$ of $\mathfrak{sl}(2)_\mathrm{R}$, and as a singlet under $\mathfrak{sl}(2)_\mathrm{L}$. Of course we could have equally well taken those other Killing spinors as part of our superalgebra.} 

Let us now also look at $G(3)$, the second superalgebra of interest in this paper. Its bosonic subalgebra is 
\begin{equation}\label{eq:g3bos}
	\mathfrak{sl}(2) \oplus \mathfrak{g}_2\,.
\end{equation}
Again we will realize $\mathfrak{sl}(2)$ as half of the $\mathfrak{sl}(2)_\mathrm{L}\oplus \mathfrak{sl}(2)_\mathrm{R}$ isometry algebra of AdS$_3$, and $\mathfrak{g}_2$ as the R-symmetry algebra. 

The maximal subgroups are this time $\mathrm{SU}(3)$, $\mathrm{SU}(2)^2$ and $\mathrm{SU}(2)$. The only choice which leads to an orbit $G/H$ of dimension $\le 7$ is $\mathrm{SU}(3)$. So our orbit is 
\begin{equation}\label{eq:g3orbit}
	G_2/\mathrm{SU}(3)\cong S^6\,.
\end{equation} 
The metric compatible with this coset turns out to be again the round one. Since its isometry group is really SO(7), it might look like we are realizing a larger symmetry group than we want. However, the coset structure on (\ref{eq:g3orbit}) implies the existence of $G_2$ invariant forms which are not invariant under SO(7).\footnote{This plays a role for example in viewing the AdS$_4\times S^6$ solution \cite{behrndt-cvetic} as a coset \cite{koerber-lust-tsimpis}.} Taking the fluxes of our solution to be proportional to these forms, we will be able to break SO(7) to $G_2$. 

The fermionic generators of $G(3)$ should now be in the $(\mathbf{2}, \mathbf{7})$ of (\ref{eq:g3bos}). This can again be achieved by (\ref{eq:eps}), this time with ${\cal N}=7$ where $I$ now transforms in the fundamental representation of $G_2$. We will see in more detail in section \ref{sub:g3spinor} how that works. 


\section{Geometrical setup} 
\label{sec:setup}

In the previous section, we have concluded that both our superalgebras can be realized with an $S^6$ symmetry orbit. We will now see more concretely how to realize this.

\subsection{Metric and fluxes} 
\label{sub:met}

We will consider metrics of the form 
\begin{equation}\label{eq:met}
	\dd s^2= e^{2A} \dd s^2_{{\rm AdS}_3} + e^{2Z} \dd z^2 + e^{2Q} \dd s^2_{S^6}\,,
\end{equation}
where $\dd s^2_{S^6}$ is the round six-sphere metric. $A$, $Q$ and the dilaton $\phi$ will be taken to be functions of $z$ only. The function $Z$ could be set to anything we like by a change of coordinates; for example, we will often consider the gauge $Z=-A$, which works nicely for AdS$_7$ solutions (see \cite[Eq.~(2.27)]{cremonesi-t}), and of course for Schwarzschild and Reissner--Nordstr\"om black holes.

The metric (\ref{eq:met}) has an SO(7) isometry group, acting on the round $S^6$. While this is the appropriate R-symmetry group for $F(4)$, for the other solutions we will consider in this paper we will need to break it to $G_2$. 

As we anticipated in section \ref{sec:ex}, the way to do that is to use fluxes. The symmetry group of a solution is defined as the group of diffeomorphisms that leave invariant not only the metric, but rather the entire set of ten-dimensional fields. This suggests that one might break the SO(7) isometry group by taking fluxes to be $G_2$ invariant forms. 

It is well-known that such forms indeed exist. One way to see this is to use the coset structure (\ref{eq:g3orbit}). (This approach is described for example in \cite[Sec.~4.4]{koerber-lust-tsimpis} in the context of AdS$_4$ solutions.) We will follow a different route. Consider a constant three-form $\hat\psi_3$ on $\mathbb{R}^7$ which is left invariant by $G_2$. (The existence of such a three-form is a possible definition of $G_2$.) Viewing $\mathbb{R}^7$ as a cone over $S^6$, and reducing $\hat\psi_3$ along the radial direction $\rho$,  
\begin{equation}\label{eq:hatpsi}
	\hat\psi_3= \rho^2 \dd \rho \wedge J + \rho^3 {\rm Re} \Omega\,, 
\end{equation}
we obtain a real two-form $J$, and a three-form ${\rm Re} \Omega$. By analogously reducing $*_7 \hat\psi_3$ we obtain another three-form ${\rm Im}  \Omega$. Together, $J$ and $\Omega$ define an SU(3) structure on $S^6$, meaning that they satisfy $J\wedge \Omega=0$, $\frac16 J^3= \frac i8 \Omega \wedge \bar \Omega$. By construction, the symmetry group of $(J, \Omega)$ is $G_2$. Moreover, $\hat\psi_3$ is closed. From these facts it follows that
\begin{equation}\label{eq:nk}
	\frac43 J^3 = i \Omega \wedge \bar \Omega \, ,\qquad J \wedge \Omega = 0 \, ,\qquad \dd J = 3 {\rm Re} \Omega \, ,\qquad \dd \Omega = -2i J^2 \,.
\end{equation}
These properties make $(J, \Omega)$ a so-called nearly-K\"ahler structure. 

We will thus use these forms as ingredients for the various field-strengths. For the NSNS three-form a priori this means
\begin{equation}\label{eq:NSNS}
 	H= h_0 {\rm vol}_{{\rm AdS}_3} + h_1 {\rm Re} \Omega + h_2 {\rm Im} \Omega  + h_3 \dd z \wedge J \,.
\end{equation}                                                                                          
In appendix \ref{app:susy} we derive from supersymmetry that in fact $h_0=0$ if the RR flux is non-zero. All the other coefficients are functions of $z$ only.  The Bianchi equation $\dd H=0$ then implies $h_2=0$, $h_1'= 3 h_3$; redefining $h_1 \equiv 3 \beta$ we find
\begin{equation}\label{BianchiNS}
	H=\dd (\beta J)\,.
\end{equation}

Using only the forms $J$ and $\Omega$ for the internal RR fluxes we write an expansion
\begin{equation}\label{eq:RR}
	F_2 = f_2 J \, ,\qquad F_4 = f_{41} J^2 + \dd z\wedge (f_{42} {\rm Re} \Omega + f_{43} {\rm Im} \Omega) \, ,\qquad
	F_6 = f_6 J^3 \, ,
\end{equation}
where again $f_2$, $f_{4i}$, $f_6$ are functions of $z$. As usual the external fluxes can be determined by duality. We will also allow $F_0\neq 0$. The Bianchi equation $\dd_H\,F=0$ implies 
\begin{equation}
\begin{split}
	&F_0=\text{const}. \, ,\qquad f_2=F_0 \beta\,,\, \\
	&f_{41}=\dfrac{1}{2}F_0 \beta^2-2\vf \, ,\qquad 
	f_{43}=\de_z \vf\, ,\qquad
	f_6=\dfrac{F_0}{6} \beta^3-2\vf \beta+\kappa\,.
\end{split}	
\end{equation}
The total RR flux can then be written more compactly as
\begin{equation}
	\label{BianchiRR}
	F\equiv F_0 + F_2 + F_4 + F_6=F_0\,e^{\beta J}+\dd_H\,(\vf\, {\rm Im} \Omega)+f_{42}\dd z \wedge {\rm Re}\Omega\,+\,\kappa J^3\,,
\end{equation}
where $\vf$ is a function of $z$, $\kappa$ is a constant, and $\dd_H \equiv \dd - H\wedge$. It is also instructive to write the $B$-transformed version as follows:
\begin{equation}\label{eq:FB}
   \tilde F\equiv e^{-B \wedge}F = F_0+\dd(\vf {\rm Im} \Omega)+f_{42}\dd z \wedge {\rm Re}  \Omega +\kappa J^3\,.
\end{equation}


\subsection{Supersymmetry} 
\label{sub:susy}

We will describe supersymmetry using generalized $G_2$ structures \cite{witt,jeschek-witt,haack-lust-martucci-t}: namely, $G_2\times G_2$ structures on the generalized tangent bundle $T \oplus T^*$. These can be parameterized by an even or odd differential form.

The supersymmetry parameters were given in (\ref{eq:eps}). The conditions for unbroken supersymmetry become then a set of spinorial differential equations for $\eta^I_{1,2}$. However, we will impose those conditions directly on one of these supercharges (in other words, we will keep only $I=1$), and will rely on other arguments to show that other supercharges must be present.

Let us then call $\eta^{I=1}_{1,2} \equiv \eta_{1,2}$. From these we can form the bispinor\footnote{One should not confuse these poly-forms $\psi_\pm$ with the constant three-form $\hat\psi_3$ in $\mathbb{R}^7$, an auxiliary entity which in (\ref{eq:hatpsi}) was instrumental in introducing the nearly-K\"ahler SU(3)-structure $(J,\Omega)$ on $S^6$.}
\begin{equation}\label{eq:psi}
	\eta_1\otimes \eta_2^\dagger\equiv  \psi_+ = i\psi_-\,.
\end{equation} 
This means that both the even form $\psi_+$ and the odd form $i \psi_-$ map to $\eta_1\otimes \eta_2^\dagger$ via the Clifford map $\dd x^{m_1}\wedge \ldots \wedge \dd x^{m_k}\mapsto \gamma^{m_1\ldots m_k}$. Indeed in odd dimensions this map is not one-to-one, and for example the forms $1$ and $\mathrm{vol}_7$ both correspond to the identity. The two polyforms are related by
\begin{equation}
	\psi_- = -* \lambda \psi_+\, ,
\end{equation}
where $\lambda C_k= (-1)^{\lfloor \frac k2 \rfloor} C_k$. 

Supersymmetry now becomes a set of equations on $\psi_\pm$ and the fluxes:
\begin{subequations}\label{eq:susy}
\begin{align}
	&\dd_H(e^{A-\phi} \psi_-) =0 \, ,\qquad \dd_H(e^{2A-\phi} \psi_+) -2 \mu e^{A-\phi} \psi_- = \frac{1}{8} e^{3A} * \lambda F \,, \label{eq:susydiff}\\
	 &(\psi_-,F)=\frac \mu2 e^{-\phi} \, ,\qquad (\psi_+,\psi_-)= \frac18 e^{2A}\,, \label{eq:susypair}
\end{align}
\end{subequations}
where the pairing $(A,B)\equiv \frac{(A\wedge \lambda(B))_7}{\mathrm{vol}_7}$.
These generalize to AdS$_3$ the Mink$_3$ system in \cite{haack-lust-martucci-t}. We have assumed here $F_0\neq 0$; the slightly more general system without this assumption, along with its  derivation from the system in \cite{10d}, is given in App.~\ref{app:susy}.\footnote{\label{foot:mmpr}The generalization of \cite{haack-lust-martucci-t} to the case with unequal spinor norms was already considered in \cite{macpherson-montero-prins}; the inclusion of the cosmological constant had already been considered by the same authors and by D.~Rosa in unpublished work.}

In (\ref{eq:f4orbits}) we saw that one way of realizing a SO(7) internal symmetry would be to use the coset $S^7\cong \mathrm{Spin}(7)/G_2$ as internal space. To analyze this possibility, we can use a logic similar to the one we used in section \ref{sub:met} for $G_2/\mathrm{SU}(3)$. Namely, we can view $\mathbb{R}^8$, the cone over $S^7$, as a Spin(7)-manifold. This means there is a Spin(7) invariant self-dual four-form $\Psi_4$. Reducing it to $S^7$ as in (\ref{eq:hatpsi}), we obtain $\Psi_4= \rho^3 \dd \rho \wedge \psi_3 + \rho^4 * \psi_3$; now the three-form $\psi_3$ defines a $G_2$ structure on $S^7$, which satisfies $\dd \psi_3 = 4 * \psi_3$ (which makes $S^7$ into a \emph{weak $G_2$} manifold). Within this Ansatz, these are the forms we can use to solve (\ref{eq:susy}); namely we need to take $\psi_+ \propto 1 + * \psi_3$, $\psi_- \propto \psi_3 + \mathrm{vol}_7$. It is immediate to see that the first equation in (\ref{eq:susydiff}) is not solved. Thus, as we anticipated after (\ref{eq:f4orbits}), the possibility $S^7\cong \mathrm{Spin}(7)/G_2$ does not work.\footnote{It is also easy to consider the same Ansatz in IIB. In that case, the Bianchi identities forbid us from using $\psi_3$ in either $H$ or $F_3$; thus the only internal flux is $F_7$, which does not break SO(8) to Spin(7). One is then led to a Freund--Rubin-type solution, which would not realize Spin(7). With some more work using the IIB version of (\ref{eq:susy}) (which is obtained by $\psi_+\leftrightarrow \psi_-$) one sees that this Ansatz does not in fact even lead to a solution.}

We then need to use the second possibility, $S^6\cong \mathrm{SO}(7)/\mathrm{SO}(6)$, as we anticipated in (\ref{eq:met}). We split the forms $\psi_\pm$ defining the generalized $G_2$ structure in longitudinal and transverse parts to $\dd z$, as in \cite{haack-lust-martucci-t}. These turn out to be pure spinors $\phi_\pm$ on $S^6$:
\begin{equation}
\begin{split}
	\psi_+&=e^Z\dd z\wedge {\rm Re} \phi_- +{\rm Re} \phi_+\,,\\
	\psi_-&=-e^Z\dd z\wedge {\rm Im} \phi_+ -{\rm Im} \phi_-\,.
\end{split}	
\end{equation}
Plugging these into (\ref{eq:susydiff}) gives rise to ``flow'' equations for $\phi_\pm$, in the sense that they determine their $z$ evolution:
\begin{subequations}\label{eq:flow}
\begin{align}
	\dd_H^6(e^{A+Z}{\rm Im} \phi_+)&=\de_z (e^A {\rm Im} \phi_-)\,, \label{eq:flowi+}\\
	\dd_H^6(e^A{\rm Im} \phi_-)&=0 \,,\label{eq:flowi-}\\
	\dd_H^6(e^{2A}{\rm Re} \phi_+)&=-2 \mu e^A {\rm Im} \phi_--\dfrac{e^{3A-Z}}8\ast_6 \lambda F_z\,, \label{eq:flowr-}\\
	\dd_H^6(e^{2A+Z}{\rm Re} \phi_-)&=\de_z(e^{2A}{\rm Re} \phi_+)+2 \mu e^{A+Z}{\rm Im} \phi_+-\dfrac{e^{3A+Z}}{8}\ast_6 \lambda F_\star \,.\label{eq:flowr+}
\end{align}	
\end{subequations}
where $\dd_H^6\equiv \dd^6 -H\wedge$ and 
\begin{equation}
	F = F_\star+\dd z \wedge F_z\,.
\end{equation}
From (\ref{eq:susypair}) we also have
\begin{equation}\label{eq:6pair}
	(F_\star,{\rm Im} \phi_+) + e^{-Z}(F_z,{\rm Im}  \phi_-)= \frac \mu2 e^{-\phi} 
	\, ,\qquad
	({\rm Re} \phi_+, {\rm Im} \phi_+) + ({\rm Re} \phi_-, {\rm Im} \phi_-) = \frac18 e^{2A}\,,
\end{equation}
where now $(A,B)\equiv \frac{(A\wedge \lambda(B))_6}{\mathrm{vol}_6}$.



\section{Solutions with $F(4)$ superalgebra} 
\label{sec:f4}

We will now present solutions with $F(4)$ superalgebra. As we anticipated, there is a unique local solution; we will first derive it in section \ref{sub:dbhat} as a near-horizon limit from a known brane solution, and then rederive it in section \ref{sub:f4susy} with the supersymmetry equations (\ref{eq:flow}), showing it is indeed locally unique. In section \ref{sub:f4glo} we will analyze its physical meaning, and we consider whether it can be cut and glued to a copy of itself by introducing D8-branes. 

\subsection{A near-horizon limit} 
\label{sub:dbhat}

We start from the solution \cite{imamura-D8,behrndt-bergshoeff-janssen}
\begin{equation}\label{eq:D2O8}
\begin{split}
	\dd s^2_{10}&= H_{\mathrm{D2}}^{-1/2} H_{\mathrm{O8}}^{-1/2}(-\dd x_0^2+\dd x_1^2) + H_{\mathrm{D2}}^{-1/2} H_{\mathrm{O8}}^{1/2} \dd x_2^2+ H_{\mathrm{D2}}^{1/2} H_{\mathrm{O8}}^{-1/2} (\dd r^2 + r^2 \dd s^2_{S^6})\,\\
	e^\phi&= H_{\mathrm{D2}}^{1/4} H_{\mathrm{O8}}^{-5/4} \, ,\qquad C_3 = \frac{H_{\mathrm{O8}}}{H_{\mathrm{D2}}} \dd x_0\wedge \dd x_1 \wedge \dd x_2\,,
\end{split}
\end{equation}
where 
\begin{equation}\label{eq:Hfun}
	H_{\mathrm{D2}}= 1+ \frac q{r^5} \, ,\qquad H_{\mathrm{O8}}= 1 + p x_2\,,
\end{equation}
and $F_0=p$. This represents a stack of D2-branes extended along directions $012$, and an O8-plane (possibly with $n_{\mathrm{D8}}<8$ D8-branes on top) along all directions except $2$.\footnote{If we considered a stack of D8-branes without an O8 (or with the O8, but with $n_{\mathrm{D8}}>8$), $H_{\mathrm{O8}}$ in (\ref{eq:Hfun}) would be replaced by a function of the form $1-p x_2$, which would stop making sense at some critical $x_2=1/p$ \cite{democratic}.} Note that while the metric and dilaton follow the ``harmonic superposition rule'' \cite{Papadopoulos:1996uq,Tseytlin:1996bh,Gauntlett:1996pb}, the three-form potential is slightly more complicated and it describes a non-homogeneous D2-brane charge distribution along the $x_{2}$ direction. In this simple case, this modification allows both the D2-branes and the O8 to be fully localized. 

We want to take a limit
\begin{equation}
	r \ll q^{1/5} \, ,\qquad  x_2 \gg 1/p\,.
\end{equation}
In other words, we want to consider the metric near the D2 but far from the O8. In this limit, the $1$ in (\ref{eq:Hfun}) can be disregarded, and the solution simplifies. Defining new coordinates $\rho$, $\alpha$ via
\begin{equation}\label{eq:rx2TOrhoalpha}
	r \equiv q^{1/3}\rho^{2/3} \cos^{-2/3}\alpha \, ,\qquad 
	x_2 \equiv p^{-1/3}\rho^{-2/3} \sin^{2/3}\alpha 
\end{equation}
and taking the small $\rho$ limit, (\ref{eq:D2O8}) becomes 
\begin{equation}\label{eq:dp}
\begin{split}
	\dd s^2_{10}&=\frac49 \left(\frac qp\right)^{1/3} \sin^{-1/3}\alpha \cos^{-5/3}\alpha \left[\dd s^2_{\mathrm{AdS_3}} + \dd\alpha^2 + \frac94 \cos^2 \alpha \dd s^2_{S^6} \right]\,\\
	e^\phi&= q^{-1/6} p^{-5/6}\cot^{5/6}\alpha \,\qquad F_0 = p \, ,\qquad F_6= *_{10} F_4= 5 q \mathrm{vol}_{S^6} \,,
\end{split}
\end{equation}
with $\rho$ having become the radial coordinate of AdS$_3$.
One can put (\ref{eq:dp}) in the gauge $Z=-A$ by defining the alternative coordinate 
\begin{equation}\label{eq:za}
	z = \tan^{2/3} \alpha\,.
\end{equation}
Indeed this turns (\ref{eq:dp}) into 
\begin{equation}\label{eq:dpz}
\begin{split}
	\dd s^2&= \frac 49\left(\frac q p\right)^{1/3} \left[  \frac{1+z^3}{\sqrt z} \dd s^2_{{\rm AdS}_3} + 
	\frac 94 \frac{\sqrt z}{1+z^3} d z^2 + \frac 94 \frac1{\sqrt z} \dd s^2_{S^6} \right]\,, \\	
	F_6 &=  5q {\rm vol}_{S^6} \, ,\qquad e^\phi = q^{-1/6} p^{-5/6} z^{-5/4}\,.
\end{split}	
\end{equation}

This solution has an SO(7) isometry acting on the $S^6$, which  also preserves the fluxes. It can also be checked directly that it enjoys ${\cal N}=8$ supersymmetry; so it is in fact $F(4)$ invariant.\footnote{\label{foot:dac}Note that this realization of $F(4)$ and the one in \protect\cite{brandhuber-oz} found in the context of $\mathrm{AdS}_{6}$ solutions differ by a choice of real section. This fact follows from their relation through a ``double analytic continuation'' that exchanges the AdS and sphere factors. For recent similar results see also \cite{Corbino:2017tfl,Dibitetto:2018gbk}.} However, rather than presenting the supersymmetry analysis directly in terms of spinors, we prefer to give it in terms of the pure spinor formalism of section \ref{sub:susy}, which will be easier to generalize in later sections.

Flux quantization is not very restrictive, given the simplicity of the fluxes and of the manifold. It restricts $2\pi F_0= 2\pi p\in \mathbb{Z}$. As for $F_6$, its only integral is on the $S^6$. (Since it does not shrink at the endpoints of the $\alpha$ interval, as we will see in section \ref{sub:f4glo}, it is to be considered a non-trivial cycle.) One has to impose that $\frac1{(2\pi)^5}\int_{S^6} F_6 \in \mathbb{Z}$; this works out to 
\begin{equation}\label{eq:fluxq}
	\frac q{6 \pi^2}\in \mathbb{Z}\,.
\end{equation}

We also see that the solution is parametrically under control: by making $q/p$ large we can make the radius large and hence the curvature small, while by making both $p$ and $q$ large we can make the string coupling $e^\phi$ small. 

Before we move to the pure spinor analysis, it is perhaps worth making a comment concerning the emergence of $F(4)$ superconformal symmetry from the near-horizon limit of the above D2-O8 system.
As discussed earlier, $F(4)$ only contains the $\mathfrak{sl}(2)_\mathrm{L}$ subalgebra of the isometries of $\mathrm{AdS}_{3}$. This hints at the fact that it should be possible to excite the right-moving degrees of freedom without further breaking supersymmetry.
This would actually correspond to turning on a large operator within the dual CFT$_2$.

On the gravity side this turns out to be realized by adding a \emph{momentum wave} to the above brane system \cite{cvetic-lu-pope-vazquezporitz}, which propagates within the common worldvolume to the D2-branes and the O8.
The explicit form of the associated warp factor entering the metric \eqref{eq:D2O8} reads
\begin{equation}
H_{\mathrm{W}}\,=\,1+Q_{\mathrm{W}}\left(x_2^3+\frac{Q_{\mathrm{D}2}}{Q_{\mathrm{O}8}}\,r^{-3}\right)\,.
\end{equation}
Upon performing the change of coordinates in \eqref{eq:rx2TOrhoalpha}, the function $H_{\mathrm{W}}$ takes the form
\begin{equation}
H_{\mathrm{W}}\,=\,1+ \frac{Q_{\mathrm{W}}}{Q_{\mathrm{O}8}}\rho^{-2}\, ,
\end{equation}
thus realizing a metric like \eqref{eq:dp} in the near-horizon limit, where empty $\mathrm{AdS}_{3}$ has been in fact replaced by a BTZ black hole with charge $Q_{\mathrm{BTZ}}=\frac{Q_{\mathrm{W}}}{Q_{\mathrm{O}8}}$.
One may check explicitly that this deformation preserves exactly the same amount of supersymmetry as the pure D2-O8 construction presented earlier.
It is worth noticing that its double analytically continued analog (see footnote \ref{foot:dac}) is the inclusion of a NUT charge in the D4-O8 system, whose effect is that of replacing the round $S^{3}$ in the transverse space by a lens space.

\subsection{Local supersymmetry analysis} 
\label{sub:f4susy}

We will now reproduce the solution (\ref{eq:dpz}) from the system (\ref{eq:flow}). 

Since we are after a solution with SO(7) symmetry, we should set to zero in (\ref{BianchiNS}), (\ref{BianchiRR}), all the flux coefficients that break it: namely,\footnote{We will see in later sections that $f_{42}=0$ is actually implied by supersymmetry anyway.}  
\begin{equation}\label{eq:fluxSO7}
	\varphi= \beta = f_{42}=0\,.
\end{equation}

Supercharges should transform in the $\mathbf{8}$ of SO(7). A way to understand spinors on $S^6$ is to work in the ambient space $\mathbb{R}^7$. The constant spinors there, 
\begin{equation}\label{eq:etaa}
	\eta_\alpha\,,
\end{equation}
indeed transform in the $\mathbf{8}$. To try and obtain more possibilities,  
we also have at our disposal functions on $\mathbb{R}^7$. We can organize these in spherical harmonics: namely, a basis is given by functions of the type
\begin{equation}\label{eq:harm7}
	s_{i_1\ldots i_k}X^{i_1}\ldots X^{i_k}
\end{equation}
where $X^i$ are the coordinates on $\mathbb{R}^7$, and $s_{i_1\ldots i_k}$ are symmetric traceless tensors. This makes the latter irreducible representations of SO(7). However, the tensor product of these with the $\mathbf{8}$ contains another $\mathbf{8}$ only if $k=1$, in which case we have $\mathbf{7}\otimes \mathbf{8}= \mathbf{56}\oplus \mathbf{8}$. Explicitly, this second possibility is given by spinors of the form 
\begin{equation}\label{eq:Xetaa}
	X^i \gamma_i \eta_\alpha\,,
\end{equation}
where $\{\gamma_i\}$ is a basis of gamma matrices in $\mathbb{R}^7$. In fact, 
\begin{equation}\label{eq:gammar}
	X_j \gamma^j = \gamma_\rho\,
\end{equation}
is the gamma matrix in the radial direction. On $S^6$ this becomes the chiral gamma; so (\ref{eq:Xetaa}) is eventually not so different from (\ref{eq:etaa}). 

As we mentioned in section \ref{sub:susy}, we will proceed by considering a single supercharge (or more precisely, a single summand in the sum over $I$ in (\ref{eq:eps})). Among the constant spinors (\ref{eq:etaa}), we have already indirectly singled out one, whose bilinears give the three-form $\hat \psi_3$; let us call it $\eta^0$. The remaining seven constant spinors $\eta^0_i$ can be obtained from $\eta^0$ as\footnote{The $i$ in (\ref{eq:eta0i}) keeps the $\eta^0_i$ Majorana: one can see this most easily in the basis where the $\gamma_i$ are purely imaginary.}
\begin{equation}\label{eq:eta0i}
	\eta^0_i = i\gamma_i \eta^0\,.
\end{equation}
If we impose supersymmetry for $\eta^0$ and SO(7) symmetry, the latter implies that the remaining $\eta_\alpha$ also solve supersymmetry. Similar to this, out of (\ref{eq:Xetaa}) we can impose supersymmetry for $X^i \gamma_i \eta^0$; SO(7) symmetry will imply supersymmetry for the other seven. 

Taking $\eta_{1,2}$ to be linear combinations of $\eta^0$ and $\gamma_\rho \eta^0$ leads to the following Ansatz for the pure spinors $\phi_\pm$ on $S^6$: 
\begin{equation}\label{eq:phi}
	\phi_+=\frac18 e^A\,e^{i\theta_+}e^{-ie^{2Q}J} \, ,\qquad
	\phi_-=\frac 18 e^A e^{i\theta_-}\,e^{3Q}\,\Omega\,,
\end{equation}
where $(J,\Omega)$ are the SU(3)-structure tensors of section \ref{sub:met}. 

We can now look at the supersymmetry conditions in (\ref{eq:flow}), with (\ref{eq:phi}) and (\ref{eq:fluxSO7}); we will work in the gauge $Z=-A$. The easiest one is (\ref{eq:flowi-}), which simply imposes $\theta_-=\frac\pi2$. From now on we will then rename $\theta_+\equiv \theta$. Next we look at (\ref{eq:flowr+}), which is also algebraic:
\begin{subequations}\label{eq:sysSO7}
\begin{equation}\label{eq:sysSO7alg}
	 6\sin\theta+4 e^{-A+Q}\mu=0\,. 
\end{equation} 
The remaining equations, (\ref{eq:flowi+}), (\ref{eq:flowr-}), contain derivatives with respect to $z$, and yield:
\begin{align}    
	&\de_z(e^{2A+3Q-\phi}) + 3 \cos\theta e^{A+2Q-\phi}=0 \, ,\\
	&\de_z(\cos\theta\,e^{3A-\phi})-6 e^{2A-6Q}\kappa +2\mu e^{A-\phi}\sin\theta=0 \, ,\\
	&\de_z(\sin\theta e^{3A+2Q-\phi})-2\mu e^{A+2Q-\phi}\cos\theta=0 \, ,\\
	&\de_z(\cos\theta\,e^{3A+4Q-\phi})+4e^{2A+3Q-\phi}+2\mu e^{A+4Q- \phi}\sin\theta=0 \, ,\\
	&\de_z(\sin\theta\,e^{3A+6Q-\phi})-F_0 e^{2A+6Q}-2 \mu e^{A+6Q-\phi}\cos\theta=0\,.
\end{align}
\end{subequations}	

One can solve (\ref{eq:sysSO7}) explicitly. By combining the equations involving $\de_z$, one can obtain one more algebraic equation beyond (\ref{eq:sysSO7alg}). Moreover, the other equations can be combined to get a decoupled differential equation that determines $Q$. Defining
\begin{equation}
	\kappa=-\dfrac{5}{6}q \, ,\qquad F_0=p \,,
\end{equation}
one eventually recovers (\ref{eq:dpz}), and no other solution.

This shows that locally (\ref{eq:dpz}) is the only solution with $F(4)$ invariance.


\subsection{Global properties} 
\label{sub:f4glo}

We are now going to investigate the physical properties of (\ref{eq:dp}) or (\ref{eq:dpz}). 

The metric is regular except at $\alpha=0$ and $\alpha=\frac \pi2$. Around $\alpha=0$ it is actually convenient to look at (\ref{eq:dpz}), which gives
\begin{equation}\label{eq:z0loc}
	\dd s^2_{10}\sim \frac1{\sqrt z} \left(\frac49 \dd s^2_{\mathrm{AdS}_3}  + \dd s^2_{S^6} \right) + \sqrt{z} \dd z^2  
	\, ,\qquad e^\phi \sim z^{-5/4}\,
\end{equation}
for the metric and dilaton. This behavior is typical of an O8 with diverging dilaton.\footnote{An O8 in flat space has a metric and dilaton $\dd s^2_{10} \sim H^{-1/2}(-\dd x_0^2 + \ldots + \dd x_8^2) + H^{1/2} \dd x_9^2$, $e^\phi \sim H^{-5/4}$ with $H= a + b|x_9|$; for $a\neq 0$ the dilaton is finite everywhere, but for $a=0$ (which is the case that reproduces (\ref{eq:z0loc})) it diverges at $z=0$.} Such a behavior has appeared in AdS$_d$ solutions for various values of $d$: for example in $d=6$ in \cite{brandhuber-oz}, for $d=7$ and $d=5$ in \cite{bah-passias-t}. In spite of the diverging dilaton, holographic computations in such cases have yielded no particular problems so far, just like in several other situations where the internal manifold of an AdS solution has a localized D-brane or O-plane.

The behavior at $\alpha=\frac\pi2$ is more problematic. We see from (\ref{eq:za}) that it corresponds to $z\to\infty$, where the metric and dilaton behave as
\begin{equation}\label{eq:zinfloc}
	\dd s^2_{10}\sim z^{5/2} \dd s^2_{\mathrm{AdS}_3} + \frac94 z^{-5/2} \left(\dd z^2 + z^2 \dd s^2_{S^6}\right) \, ,\qquad
	e^\phi  \sim z^{-5/4}\,.
\end{equation}
This looks like the behavior of a D2-brane metric $H_\mathrm{D2}^{-1/2} (-\dd x_0^2 + \dd x_1^2 + \dd x_2^2)+ H_\mathrm{D2}^{1/2} (\dd x_3^2 + \ldots + \dd x_9^2)$ near the D2, with $z$ being interpreted as the radial variable in the $\mathbb{R}^7_{3\ldots 9}$ transverse to it. However, from the standard D2 metric in flat space  one actually would obtain (\ref{eq:zinfloc}) at \emph{small} $z$, not at large $z$ (which is the domain where we obtained it). So in fact (\ref{eq:zinfloc}) is the behavior \emph{far} from a D2-brane in flat space, in a situation where one takes $H_\mathrm{D2}\propto r^{-5}$ rather than the standard $H_\mathrm{D2} = 1 + \frac{r_0^5}{r^5}$. 

Related to this, the internal space is non-compact: indeed $z=\infty$ is at infinite distance, since $\int z^{-1/4}\dd z$ diverges. A perhaps more satisfying physical measure of this non-compactness is the dual CFT$_2$ central charge. Standard computations \cite{henningson-skenderis} adapted to the case with non-constant dilaton and warping (see e.g.~\cite[Sec.~4.1]{cremonesi-t} for a discussion) give
\begin{equation}\label{eq:c}
	c \propto \int_{M_6} e^{A-2\phi} \mathrm{vol}_{M_7} = 2 p^{1/3}q^{5/3}\int_0^\infty z \dd z \,.
\end{equation}                                                                                  
Thus the dual CFT would have infinitely many degrees of freedom. Notice that this divergence has nothing to do with the O8, which is located at $z=0$. (For AdS$_7$ solutions with an O8 with diverging dilaton, the central charge $a_\mathrm{6d}$ is finite and matches a field theory computation \cite{bah-passias-t}.)    

In light of recent AdS solutions with D8-branes such as \cite{afrt,10letter,cremonesi-t}, it is natural to try to get rid of this non-compactness by cutting the solution at some value $z=z_0$, and gluing it to a second copy of it. For simplicity, let us try to glue two identical copies. Namely, let us use (\ref{eq:dpz}) for $z\in[0,z_0]$, and for $z\in[z_0,2z_0]$ the same solution (\ref{eq:dpz}) but with $z \to 2z_0-z$. At the gluing locus $z=z_0$, there is a defect; to interpret it as a D8, $F_0$ should jump. Again for simplicity, let us assume that its value goes $p$ to $-p$ as we cross the D8. Looking at the metric in (\ref{eq:dpz}), we see that the sign of $q$ must also flip.

With this choice, the metric and dilaton are continuous, and the fluxes have a discontinuity
\begin{equation}\label{eq:jumpF}
	\Delta F_0 = -2 F_0 = -2p \, ,\qquad \Delta F_6 = -2 F_6 = -10 q \mathrm{vol}_6\,.
\end{equation}  
The Bianchi identities in presence of a D8-brane stack at $z=z_0$ are
\begin{equation}\label{eq:bianchi}
	dF = \frac1{2\pi}n_{\mathrm{D8}} e^{2\pi f} \delta(z-z_0)\,,
\end{equation}
where $f$ is the gauge flux on the D8 worldsheet.  Comparing (\ref{eq:jumpF}) and (\ref{eq:bianchi}) we see that  
\begin{equation}\label{eq:jumppq}
	p=-\frac{n_\mathrm{D8}}{4\pi} \, ,\qquad\mathrm{Tr} f= \mathrm{Tr}f\wedge f = 0   \, ,\qquad \mathrm{Tr}f\wedge f \wedge f  = 5\frac qp \mathrm{vol}_6\,.
\end{equation} 
This worldsheet flux means that the D8 is in fact a D8/D2 bound state. 
                 
It is a bit surprising that in (\ref{eq:jumppq}) the position $z_0$ is not fixed by the Bianchi identities. From the point of view of the equations of motion, the jump in the derivatives of the functions in the metric is realated to the D8 data. Since these jumps depend on our choice of $z_0$, one would imagine that they should fix the position of the D8. On the other hand, supersymmetry and the Bianchi identities imply the equations of motion, so whatever condition comes from the equations of motion should follow from (\ref{eq:jumppq}) already. Checking this explicitly is hard because of the subtleties involved in the non-Abelian DBI action, but a preliminary computation makes it plausible.

For a complete analysis, one should also check that supersymmetry is preserved on the D8. This leads to checking certain $\kappa$-symmetry conditions, that can be formulated in terms of pure spinors along the lines of \cite{martucci-smyth}. This leads to imposing either that the pull-back of $e^f {\rm Im} \phi_+$ and $X\cdot e^f \phi_-$ vanish (for any $X$ section of $T \oplus T^*$), or that the pullback of $e^f {\rm Re} \phi_+$ reproduces the DBI action. This would lead to a version of the equations in \cite{marino-minasian-moore-strominger}, but again the non-Abelian nature of the equations makes an explicit check difficult. In the similar but simpler situation for D8/D6 in AdS$_7$ solutions \cite{afrt}, this $\kappa$-symmetry condition was automatically satisfied.

If we try to proceed with a physical interpretation in spite of these difficulties, the central charge of the dual theory would be just the same as (\ref{eq:c}), but with the integral $\int z \dd z$ now cut off at a finite value of $z$. Intriguingly, the prefactor $p^{1/3} q^{5/3}$ in (\ref{eq:c}) matches the behavior of the three-sphere partition function or the thermal free-energy coefficient found in \cite{ajtz,jafferis-klebanov-pufu-safdi,guarino-jafferis-varela} for AdS$_4$ solutions with Romans mass.

As we mentioned, the procedure of gluing two pieces of the solution (\ref{eq:dp}) is partially inspired by AdS$_7$ solutions with D8-branes \cite{afrt,10letter,cremonesi-t}. An important difference with that case is that the metric we are starting from only depends on the ratio $q/p$, while for AdS$_7$ the local metric depends on two parameters; this presumably would prevent extending the construction above to more than one D8 stack (although we have not tried systematically to do this). Another context that bears some resemblance to our case is non-Abelian T-duality, which can be used to generate AdS solutions which have a non-compact internal space (see for example \cite{kelekci-lozano-macpherson-ocolgain} for an AdS$_3$ example). In that case, one often manages to interpret the dual field theory as an infinite quiver; cutting the solution and gluing it to a copy of itself, as we have done here, corresponds to an intuitively similar cutting and gluing procedure on the quiver --- see for example \cite[Fig.~7,8]{lozano-nunez} for an AdS$_5$ example of this.



\section{Solutions with $G(3)$ superalgebra} 
\label{sec:g3}

We will now look for solutions with $G(3)$ superalgebra. In this case, the R-symmetry is $G_2$; it will be realized as the isometry group SO(7) of $S^6$ broken by the fluxes, as described in section \ref{sub:met}. 

We will start in section \ref{sub:g3spinor} by finding the appropriate supersymmetry parameters; this 
requires a little more work than in section \ref{sub:f4susy}. In section \ref{sub:g3sys} we will then apply the resulting spinorial Ansatz to the supersymmetry conditions in (\ref{eq:susy}). 

\subsection{Internal spinors} 
\label{sub:g3spinor}

We need spinors transforming in the $\mathbf{7}$ of $G_2$ on $S^6$. 
We will again consider $S^6$ as embedded in $\mathbb{R}^7$ with coordinates $X^i$, as in section \ref{sub:f4susy}. As we saw there, of the 8 independent constant Majorana spinors $\eta_\alpha$ on $\mathbb{R}^7$, one is $G_2$ invariant, and we called it $\eta^0$; the others, which can be written as in (\ref{eq:eta0i}), transform in the $\mathbf{7}$. 

Again we also have at our disposal functions on $\mathbb{R}^7$, for which (\ref{eq:harm7}) is a basis. The symmetric traceless tensors $s_{i_1\ldots i_k}$ that appear there, form irreducible representations of SO(7); it turns out that they are also irreducible as representations of $G_2\subset$ SO(7).

So in principle we have to combine the symmetric traceless tensor representations in (\ref{eq:harm7}) with the $\eta^0_i$ and with $\eta^0$ to obtain a $\mathbf{7}$. Combining them with the singlet $\eta^0$ only gives rise to the combination 
\begin{equation}\label{eq:Xieta}
	X^i \eta^0\,.
\end{equation}                  
The only symmetric traceless tensors (\ref{eq:harm7}) whose tensor products with a $\mathbf{7}$ contain another $\mathbf{7}$ are the ones with one and two indices:
\begin{equation}
	X_i \, ,\qquad X_i X_j-\frac17\delta_{ij}\,;
\end{equation}
namely, respectively a $\mathbf{7}$ and a $\mathbf{27}$. The way to combine them with the $\eta^0_i = \gamma_i \eta^0$ to give a $\mathbf{7}$ is respectively by contracting with the $G_2$ invariant tensor $\phi_{ijk}$, and by direct contraction:
\begin{equation}\label{eq:714}
	\phi_{ijk} X_j \gamma_k \eta^0 \, ,\qquad \left(X_i X_j-\frac17\delta_{ij}\right) \gamma^j \eta^0=  
	\left(X_i X_j \gamma^j - \frac17 X_i\right) \eta^0\,. 
\end{equation}
In fact the first in (\ref{eq:714}) can be rewritten using (see e.g.~\cite[(3.8)]{kmt})
\begin{equation}\label{eq:gijeta}
	\gamma_{ij} \eta^0 = i \phi_{ijk} \gamma^k \eta^0\,.
\end{equation}
Putting together (\ref{eq:eta0i}), (\ref{eq:Xieta}), (\ref{eq:714}) and (\ref{eq:gijeta}) we conclude that the most general set of spinors on $\mathbb{R}^7$ that transform in the $\mathbf{7}$ of $G_2$ can be written as 
\begin{equation}\label{eq:g3spinor}
	\eta_i = (a X_i + i b \gamma_i) \eta^0 + (c X_i + i d \gamma_i) X_j \gamma^j \eta^0\,.
\end{equation}
Recall that $X_j \gamma^j$ in (\ref{eq:g3spinor}) is just the radial gamma matrix, (\ref{eq:gammar}); this will simplify the analysis.

In (\ref{eq:eps}), we need two sets of internal spinors, $\eta^I_1$ and $\eta^I_2$. The index $I$ should in this case be identified with the index $i$ in (\ref{eq:g3spinor}): indeed the solution will have ${\cal N}=7$ supersymmetry. We can use the decomposition (\ref{eq:g3spinor}) for both $\eta^I_1$ and $\eta^I_2$, with two sets of parameters $a_a,\,\ldots,d_a$, $a=1,2$. Moreover, as we discussed in section \ref{sub:susy}, we can impose the supersymmetry conditions on one particular choice of $I$, say $I=1$; R-symmetry then implies that they are also satisfied for the remaining values of $I$. So our internal spinors read
\begin{equation}\label{eq:g3spinors}
	 \eta_a =  (a_a X_1 + i b_a \gamma_1) \eta^0 + (c_a X_1 + i d_a \gamma_1) \gamma_\rho \eta^0\,. 
\end{equation}
We can see already now that the case $a_a=c_a=0$ leads to the $F(4)$ case we considered in section \ref{sec:f4}. While in section \ref{sub:f4susy} we took the internal spinors to be a linear combination of $\eta^0$ and $\gamma_\rho \eta^0$, we could in fact have chosen any other element in the $\textbf{8}$ of SO(7) taken from (\ref{eq:etaa}) and (\ref{eq:Xetaa}); the case $a_a=c_a=0$ in (\ref{eq:g3spinors}) is indeed of that type. (Recall that SO(7) mixes the $\eta^0_i$ with $\eta^0$, while $G_2$ does not.)

To compute the pure spinors on $S^6$ corresponding to (\ref{eq:g3spinors}), it is convenient to first compute the bilinears $\eta_1 \eta_2^\dagger$ as a poly-form $\hat \psi=\hat \psi_+ + \hat \psi_-$ on $\mathbb{R}^7$ with $\hat\psi=\hat\psi_+=i\hat\psi_-$ as in (\ref{eq:psi}), and then reduce it to $S^6$ as 
\begin{equation}
\begin{split}
	\hat\psi|_{\rho=1}\,&=\,\dd \rho\wedge {\rm Re} \,\phi_-\,+\, {\rm Re} \phi_+\,,\\
	-i\hat\psi|_{\rho=1}\,&=\dd \rho\wedge {\rm Im} \,\phi_+\,+\, {\rm Im} \phi_-\,.
\end{split}	
\end{equation}
(\ref{eq:hatpsi}) can be now recovered as a particular case of the above relations, where $\eta_2=\eta_1$, in which case $\hat\psi_- = \hat\psi_3 + \mathrm{vol}_7$.


\subsection{System} 
\label{sub:g3sys}

We can now plug the $\phi_\pm$ obtained with the spinors (\ref{eq:g3spinors}) in (\ref{eq:flow}). The resulting system of equations is much more complicated than the one we had for $F(4)$, and we will not write it down here. With some effort, however, it can be solved; we will describe here the main steps. 

First of all, some of the equations are algebraic (i.e.~they do not involve $\de_z$), similar to (\ref{eq:sysSO7alg}) for $F(4)$. These can be completely solved for the functions $a_a,\,\ldots,\,d_a$ appearing in the spinors (\ref{eq:g3spinors}),
\begin{equation}
\begin{split}
&a_1=-\dfrac{2(\a- \b)\t}{b_2 (1+\a^2)} \, ,\qquad
b_1 = \dfrac{\t}{b_2} \, ,\qquad
c_1 =-\dfrac{2 \a(\a- \b)\t}{b_2 (1+ \a^2)} \, ,\qquad
d_1 = \dfrac{ \b \t}{b_2}\,,\\
&a_2 = \dfrac{2b_2(\a-\b)}{1+\a^2} \, ,\qquad
c_2 = -\dfrac{2b_2 \a( \a- \b)}{1+\a^2}\,,\qquad
d_2 = -b_2 \b\,,
\end{split}	
\end{equation}
 and for
\begin{equation}  
\begin{split}
e^A\,&=\, 8(1+ \b^2)\t\,,\\
e^Q\,&=\,\dfrac{32(-\a+\b)(1+ \a \b)(1+\b^2)\t}{\sqrt{(1+\a^2)(1+\b^2)(1-16\a\b+9 \b^2+\a^2(9+\b^2))}}\,,\\
\tan\theta\,&=\, \dfrac{4\b+4\a^2\b-2\a(1+\b^2)}{-1+3\b^2+\a^2(-3+\b^2)}\,.
\end{split}
\end{equation}  
Some more algebraic conditions can be found by linear combinations. They can be used to fix the remaining fluxes, whose expressions we will see later.

The remaining differential equations can be simplified by imposing consistency with the Bianchi identities, but they still look daunting:
\begin{subequations}
\begin{align}
	\label{eq:phi'}
	&\phi'\,=\,\dfrac{e^{Z+\phi}F_0\Delta_1(-1+3\b^2+\a^2(-3+\b^2))(5+16\a\b-3\b^2+\a^2(-3+5\b^2))}{8\,(-\a+\b)(1+\a\b)^3} \,,\\
	\label{eq:ab'}
	&\dfrac{4\b'}{1+\b^2}-\dfrac{4\a'}{1+\a^2}\,=\,\dfrac{3 e^{Z+\phi}F_0\Delta_2(-1+3\b^2+\a^2(-3+\b^2))}{(1+\a\b)^2}\,,	
\end{align}	
\end{subequations}
where
\begin{equation}  
\begin{split}
\Delta_1&=\sqrt{1+\dfrac{4(\a-2(1+\a^2)\b+\a\b^2)^2}{(-1+3\b^2+\a^2(-3+\b^2))^2}}\,,\\
\Delta_2&=\sqrt{\dfrac{(1+\a^2)(1+\b^2)(1-16 \a\b+9\b^2+\a^2(9+\b^2))}{(-1+3\b^2+\a^2(-3+\b^2))^2}}\,,
\end{split}
\end{equation}  
and a prime denotes differentiation with respect to $z$.
The equation (\ref{eq:phi'}) just fixes the dilaton, so we can solve at the end. The relevant equation is (\ref{eq:ab'}). We can still use the gauge freedom in selecting $Z$ (see discussion after (\ref{eq:met})) to simplify it. A good choice is
\begin{equation}  
e^Z= \dfrac{4(1+\a\b)^2}{3F_0\sqrt{(1+\a^2)(1+\b^2)(1-16\a\b+9\b^2+\a^2(9+\b^2))}}\,.
\end{equation}  
The equation \eqref{eq:ab'} now becomes solvable:
\begin{equation}  
\label{AlfaOfBeta}
\dfrac{\b'}{1+\b^2}=-1+\dfrac{\a'}{1+\a^2}\,\qquad \Rightarrow\qquad \a=\dfrac{\tan z+\b}{1-\b \tan z}\,.
\end{equation}  
It might look as though every choice of $\b$ leads to a different solution. In fact, however, upon substituting the solution \eqref{AlfaOfBeta}, $\b$ disappears from all the physical fields, and there is a unique solution. After a change of coordinates $z=\arctan y$, the dilaton and the metric coefficients in (\ref{eq:met}) read:
\begin{subequations}\label{eq:g3tot}
\begin{equation}\label{eq:g3sol}  
\begin{split}
\dd s^2 &=  e^{2A} \dd s^2_{\mathrm{AdS}_3} + e^{2Y} \dd y^2 + e^{2Q} \dd s^2_{S^6} \,\\ 
e^{2A}\,&=\,\left(\dfrac{q}{p}\right)^{1/3}\,\dfrac{(1+9\,y^2)\,(1+y^2)^{1/3}}{4\,y^{5/3}} \,\\
e^{2Q}\,&=\,\left(\dfrac{q}{p}\right)^{1/3}\,\dfrac{y^{1/3}}{(1+y^2)^{2/3}}\,,\\
e^{2Y}\,&=\,\left(\dfrac{q}{p}\right)^{1/3}\,\dfrac{4}{9\,(1+9\,y^2)(1+y^2)^{5/3}\,y^{5/3}}\,,\\
e^\phi&=\,\dfrac{y^{5/6}}{q^{1/6}\,p^{5/6}\,(1+y^2)^{2/3}}\,.
\end{split}
\end{equation}  
The fluxes read, in terms of (\ref{eq:FB}) and (\ref{BianchiNS}):
\begin{equation}\label{eq:g3flux}  
\begin{split}
&F_0=2p \, ,\qquad \vf= -\left(\dfrac{p}{q}\right)^{1/3}\dfrac{q\,y^{2/3}}{(1+y^2)^{1/3}} \, ,\qquad f_{42}=0 
\, ,\qquad\\
&\kappa = -\frac 56 q \, ,\qquad 
\beta =\left(\dfrac{q}{p}\right)^{1/3}\dfrac{y^{4/3}}{(1+y^2)^{2/3}}\,.
\end{split}
\end{equation}  
\end{subequations}

Flux quantization for this solution might look more complicated, given that more fluxes are switched on; but in fact the $\dd$-closed fluxes (\ref{eq:FB}) are rather simple, given that $f_{42}=0$. Moreover the term $\dd(\varphi {\rm Im} \Omega)$ is exact and gives no contribution. So in fact one ends up with (\ref{eq:fluxq}) again. The solution is also parametrically under control, for reasons mentioned below (\ref{eq:fluxq}).

The solution in (\ref{eq:g3sol}) presents the same issues as the $F(4)$ solution of section \ref{sec:f4}. Namely, one can see that there are only two singularities, at $z=0$ and $z=\infty$, where the local behavior is respectively (\ref{eq:z0loc}) and (\ref{eq:zinfloc}) again. 

One might again try to glue the solution to a copy of itself, as we sketched in section \ref{sub:f4glo} for $F(4)$ solutions. The problem is very similar, but a little less restrictive because $F_4$ in (\ref{eq:g3flux}) is no longer zero; because of this, in the analogue of (\ref{eq:jumppq}) one need not require $\mathrm{Tr}(f\wedge f)=0$.



\section{${\cal N}=1$ solutions with $G_2$ flavor symmetry} 
\label{sec:g2}

We will now consider solutions which have only ${\cal N}=1$ supersymmetry, but also $G_2$ flavor symmetry. 

This is in a sense a mix of the two previous sections. As in section \ref{sec:g3}, we are going to allow all components in (\ref{BianchiRR}) and (\ref{BianchiNS}), so that the SO(7) isometry group of the $S^6$ is broken to $G_2$. The internal spinors will be taken however equal to those of section \ref{sec:f4}, namely a linear combination of the $G_2$ invariant $\eta^0$ and of $\gamma_\rho \eta^0$. Since the internal symmetry is only $G_2$ and not SO(7), it cannot be used to generate new supercharges as in section \ref{sec:f4}. 

We will set up the supersymmetry system in section \ref{sub:g2susy}, and then proceed in later sections to study it. We will obtain some numerical solutions with O2- and O8-planes. 

\subsection{System} 
\label{sub:g2susy}

Thus we start again with the system (\ref{eq:flow}), and the pure spinors in (\ref{eq:phi}). Unlike in section \ref{sub:f4susy}, we allow for all possible components in the fluxes (\ref{BianchiRR}) and (\ref{BianchiNS}), but still our analysis will parallel the one there. We also go back to the gauge $Z=-A$.

Again (\ref{eq:flowi-}) fixes $\theta_-=0$, and from now on $\theta_+ \equiv \theta$. The other equations can be rewritten in a more compact way if one defines
\begin{equation}
	\tau= \beta +ie^{2Q} \, ,\qquad \xi \equiv  \beta +i \alpha\, \,,
\end{equation}
with $\alpha$ defined by 
\begin{equation}
	\varphi \equiv \dfrac{F_0}{4} \alpha^2\,.
\end{equation}
The expression (\ref{eq:flowr+}) now reads
\begin{subequations}\label{eq:G2sys}
\begin{equation}
\label{algebraic}
f_{42}=0 \, ,\qquad \dfrac{e^{2A}F_0}{4}\de_z(\alpha^2)-2 \mu e^{3Q-\phi}\,
+3e^{A-\phi}{\rm Re} [\tau e^{i\theta}]=0
\end{equation}
while (\ref{eq:flowi+}) gives the differential equation
\begin{equation}
\label{diffSimple}
3 {\rm Im} [\tau e^{i\theta}]e^{A-\phi} +  \de_z(e^{3Q+2A-\phi})=0\,.
\end{equation}
Then, (\ref{eq:flowr-}) gives
\begin{equation}
\begin{split}
\label{system1}
&\de_z(e^{3A-\phi}\cos \theta)+2\mu e^{A-\phi}\sin\theta-e^{2A-6Q}(F_0 {\rm Re} [\xi^3]+6\kappa)=0 \,,\\
&\de_z(e^{3A-\phi+2Q}\sin\theta)-2\mu e^{A-\phi+2Q}\cos\theta - e^{3A-\phi} \cos\theta\de_z\beta+e^{2A-2Q}F_0 {\rm Re} [\xi^2]=0\,,\\
&\de_z(e^{3A-\phi+4Q}\cos\theta)+2\mu e^{A-\phi+4Q}\sin\theta +4 e^{2A-\phi+3Q}+2 e^{3A-\phi+2Q}\sin\theta\de_{z}\beta+e^{2A+2Q}F_0 \beta = 0\,,\\
&\de_z(e^{3A-\phi+6Q}\sin\theta)-2\mu e^{A-\phi+6Q}\cos\theta+12 e^{2A-\phi+3Q}\beta -3 e^{3A-\phi+4Q}\cos\theta\de_z\beta-e^{2A+6Q}F_0 =0\,.
\end{split}
\end{equation}
Finally, one of the pairing equations (\ref{eq:6pair}) is already satisfied, while the other gives
\begin{equation}\label{pairing}
\im\left[(F_0 \tau^3-12\vf \tau+6\kappa)e^{i\theta}\right]+F_0\,e^{3Q+A} \de_z\alpha^2-4\mu e^{6Q-A-\phi}=0\,.
\end{equation}
\end{subequations}

The system (\ref{eq:G2sys}) has a useful symmetry: it is invariant under the rescaling
\begin{equation}
\begin{split}
	&\kappa \to a \kappa \, ,\qquad F_0 \to b F_0 \, ,\\
	&e^A\to \left(\frac ab\right)^{1/6}e^A \, ,\qquad
	e^Q\to \left(\frac ab\right)^{1/6}e^Q \, ,\qquad
	e^\phi\to (a b^5)^{-1/6} e^\phi \, ,\\
	&z\to \left(\frac ab \right)^{1/3} z \, ,\qquad
	\beta \to \left(\frac ab\right)^{1/3} \beta \, ,\qquad
	\varphi \to \frac {a^{2/3}}{b^{1/3}} \varphi \, ,
\end{split}	
\end{equation}
where $a$ and $b$ are constant. This rescaling is inspired by the $G(3)$ solution (\ref{eq:g3tot}). Thanks to this symmetry, once we find a solution we can rescale it by increasing $a$ and $b$, to make the curvature and string coupling arbitrarily small.


\subsection{Local analysis} 
\label{sub:g2loc}

We could not solve the system (\ref{eq:G2sys}) analytically; we thus decided to look for numerical solutions. In preparation to that, in this subsection we will solve the system perturbatively around some notable loci which can be used as endpoints for the $z$ interval. These results will then be used as boundary conditions for the numerical study. Notice that the system (\ref{eq:G2sys}) is autonomous (i.e.~$z$ does not appear explicitly), so we will expand around $z=0$ without loss of generality. 

First we consider what happens around a point where the $S^6$ shrinks smoothly. This happens if 
\begin{equation}\label{eq:reg}
	e^{-2A}\sim e^{-2A_0} \, ,\qquad e^{2Q}\sim e^{-2A_0} z^2\,,
\end{equation}
for a constant parameter $A_0$. Indeed in this case (\ref{eq:met}) (with $Z=-A$) is proportional to $\dd s^2_{\mathrm{AdS}_3}+ \dd s^2_{\mathbb{R}^7}$, with the $\mathbb{R}^7$ in polar coordinates. All the other functions are simply assumed to have a regular Taylor expansion. Imposing (\ref{eq:reg}) in (\ref{eq:G2sys}) and solving perturbatively, we get $\kappa=0$ and 
\begin{equation}\label{eq:regexp}
\begin{split}
	&e^A= e^{A_0}-\frac{e^{-3A_0}}{49}(5\mu^2 + f_0 \mu-f_0^2) z^2+ O(z^4) 
	\, ,\\
	& e^Q = e^{-A_0} z + \frac{e^{-5A_0}}{882}(42 \mu^2 -2f_0 \mu + f_0^2) z^3 + O(z^5) \,,\\
	& e^\phi = e^{\phi_0} + \frac{e^{-4A_0 + \phi_0}}{98}(2 \mu^2 -6 f_0 \mu + 7 f_0^2) z^2+ O(z^4) \, ,\\ 
	&\beta = \frac2{21}e^{-4A_0}(\mu+ f_0)z^3 + O(z^5) \,,\\
	& \vf=\frac{e^{-5A_0-\phi_0}}{28}(4 \mu - f_0)z^4 + O(z^6) \, ,\\ 
	&\theta= \pi + \frac17 e^{-2A_0}(4 \mu + f_0) z + O(z^3)\,.
\end{split}	
\end{equation}
where $f_0 \equiv F_0 e^{A_0+ \phi_0}$, and $\phi_0$ is a constant parameter. 

As we will see in the next section, evolving from (\ref{eq:regexp}) one finds solutions which behave like O8- and O2-planes. Moreover, both the $F(4)$ and $G(3)$ solutions have O8 singularities. Using these solutions as inspiration, we were able to find local solutions corresponding to both these objects. In both cases, the leading behavior can be inferred from the corresponding solutions in flat space; it is less trivial to guess the subleading powers. (A similar problem was overcome in \cite[(5.6)--(5.7)]{rota-t} for D6 and O6 singularities.) For the O8 the expansion reads 
\begin{equation}\label{eq:O8exp}
\begin{split}
	&e^A = \frac 2{c^3} z^{-1/4} - \frac34 c z^{3/4} + O(z^{7/4}) \, ,\qquad
	e^Q = -\frac 2{c^3} z^{-1/4} - \frac14 c z^{3/4} + O(z^{7/4})\,,\\ 
	&e^\phi = \frac1{F_0}\left(-\frac 2{c^3} z^{-5/4} + \frac34 c z^{-1/4}+ O(z^{3/4})\right)
	\, ,\qquad
	\beta = 3 z - \frac{c^4}2 z^2 + O(z^4) \,, \\ 
	&\vf= F_0 \left(\frac 2{c^8}- \frac12z^2 + O(z^4)\right) \, ,\qquad \theta= \frac\pi2-\frac14 c^6 z^{3/2} \,.
\end{split}	
\end{equation}
For simplicity we have set $\kappa=0$ here. For the O2 we find 
\begin{equation}\label{eq:O2exp}
\begin{split}
	&e^A = a_0 z^{-1/4}\left(1+\frac 3{4 a_0 q_0} z+ O(z^2)\right) \, ,\qquad
	e^Q = q_0 z^{1/4}\left( 1 + \frac1{4a_0 q_0} z+ O(z^2)\right)\,,\\ 
	&e^\phi = \frac{q_0^6 a_0}{6\kappa}z^{1/4}\left(1-\frac 3{4 a_0 q_0} z + O(z^2)\right)
	\, ,\qquad
	\beta = z^2\left(\frac{q_0^2}{a_0^2}+ \frac{F_0 q_0^8}{12\kappa}\right) + O(z^3)\,,\\
	&\vf= \frac{6 \kappa}{ a_0^3 q_0^3}z^2- \left(\frac{18 \kappa}{a_0^4q_0^4}+ \frac{F_0 q^2_0}{2 a_0^2}\right)	+ O(z^2) \, ,\qquad \theta = -z^{3/2} \left(\frac4{a_0^2}+\frac{F_0 q_0^6}{6 \kappa}\right)+ O(z^2)\,.
\end{split}	
\end{equation}
In the above, $c$, $a_0$ and $q_0$ are constant parameters.


\subsection{Some numerical solutions} 
\label{sub:g2num}

As is often done, we used the perturbative expansions we just saw (evaluated at small values of $z$) as initial conditions for a numerical evolution of the system (\ref{eq:G2sys}). In this section we give some sample solutions we found this way. 

\paragraph{Single O8.} We will first describe a solution with a single O8. We obtained it by evaluating at small $z$ the expansion (\ref{eq:regexp}) around a regular point, and using it as an initial condition for a numerical evolution. 

\begin{figure}[h]
\centering
\includegraphics[width=7cm]{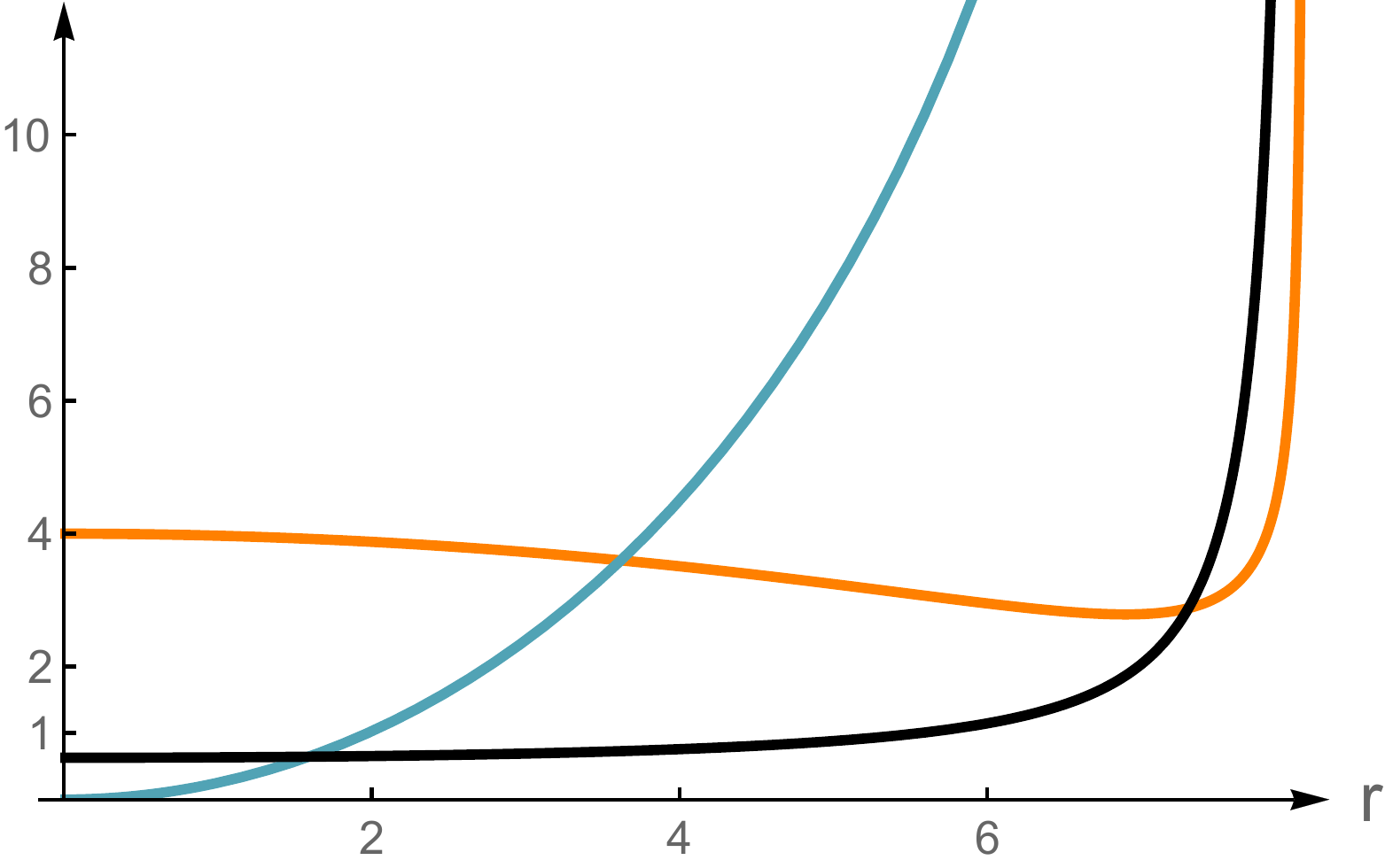}
\caption{A solution with a single O8 (at the right extremum). In orange $e^{2A}$, in blue $e^{2Q}$ and in black the dilaton $e^{\phi}$.}
\label{fig:O8}
\end{figure}

We selected a particular class, namely we assumed $\beta=\kappa=0$. This implies in particular
\begin{equation}
H=0 \, ,\qquad F=F_0+ \dd(\varphi {\rm Im}  \Omega )\,.
\end{equation}
In this case, the system (\ref{eq:G2sys}) can be reduced. In particular, it can be shown that $A$, $\phi$, $\varphi$ are determined as functions of $Q$ and $\theta$:
\begin{equation}
\label{kbFunctions}
e^A= \dfrac{-1}{\cos\theta Q'} e^{-Q} \, ,\qquad \varphi= \dfrac{F_0}{4}e^{4Q}-\dfrac{F_0 \tan\theta}{\mu Q'}e^{2Q} \, ,\qquad e^\phi= \dfrac{- \mu Q'}{F_0} e^Q\,.
\end{equation}
At this point, \eqref{system1} reduces to two differential equations for $Q$ and $\theta$:
\begin{equation}
\begin{split}
	\label{kbDiff}
	&Q''= - (\cos\theta Q')^2+\dfrac{2}{3} \tan\theta Q' \theta'\,,\\
	&\theta' =3 \cos^2\theta Q'(-2\tan\theta+e^{2Q}\mu Q')\,.	
\end{split}
\end{equation}
A numerical study of this reduced system produces solutions such as the one in Fig.~\ref{fig:O8}.

Around the right endpoint $z=z_0$, we checked that the behavior is the one in (\ref{eq:O8exp}), which produces an O8 solution:
\begin{equation}
\dd s^2\approx r^{-1/2}(\dd s^2_{\mathrm{AdS}_3}+\dd s^2_{S^6})+ r^{1/2} \dd z^2\,\, ,\qquad e^\phi \approx r^{-5/4}\,,
\end{equation}
where $r=z_0-z$.

\begin{figure}[h]
\centering
\includegraphics[width=7cm]{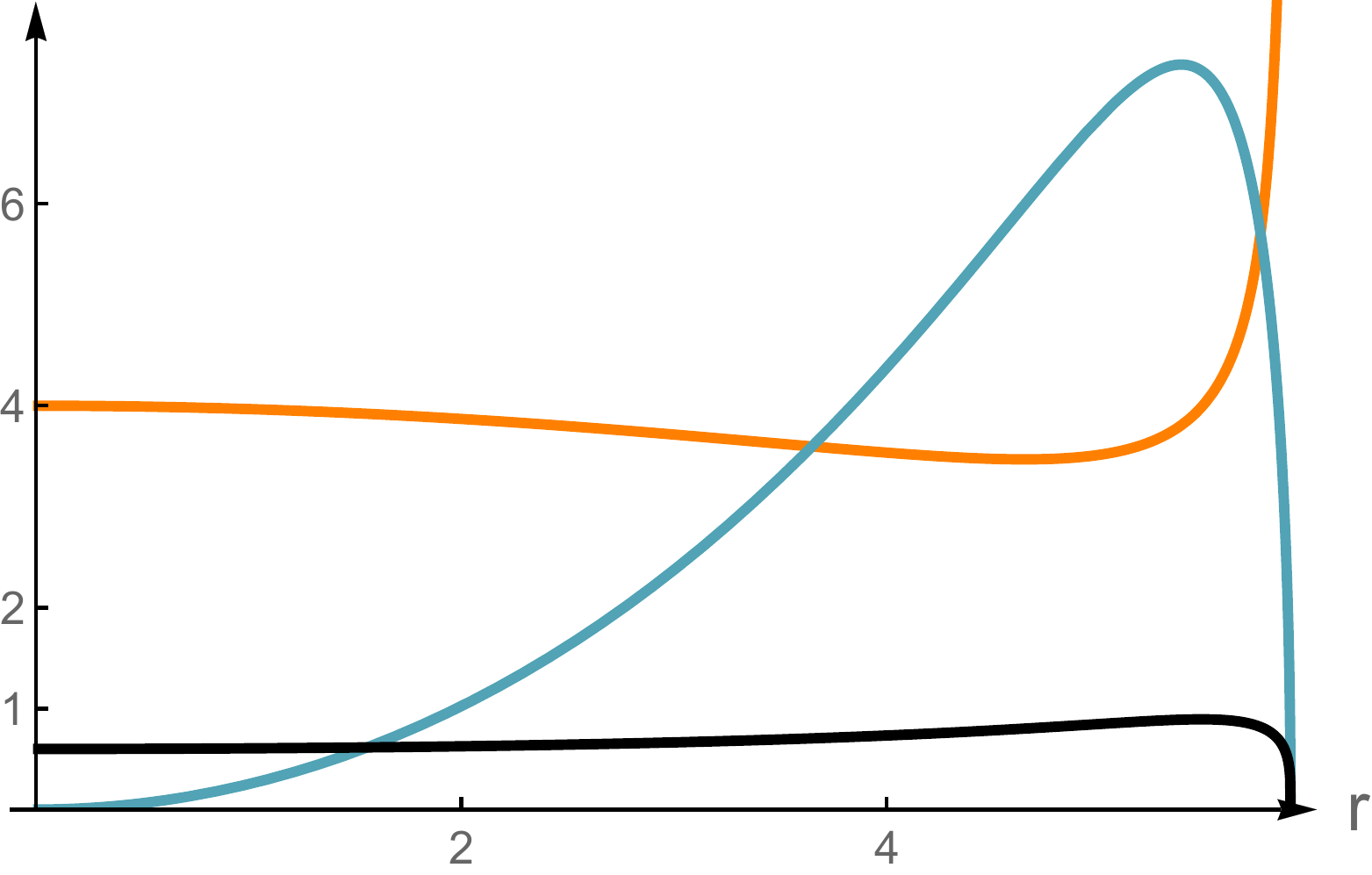}
\caption{A solution with a single O2 (at the right extremum). In orange $e^{2A}$, in blue $e^{2Q}$ and in black the dilaton $e^{\phi}$.}
\label{fig:O2}
\end{figure}

\paragraph{Single O2.}
We now again evolve numerically from (\ref{eq:regexp}), but keeping all the flux components in (\ref{BianchiRR}) and (\ref{BianchiNS}). We generically find a solution such as the one in Fig.~\ref{fig:O2}.

Around the right extremum $z=z_0$, this time the behavior is the one in (\ref{eq:O2exp}), which results in an O2:
\begin{equation}
\dd s^2\approx r^{-1/2}\dd s^2_{\mathrm{AdS}_3}+ r^{1/2} (\dd z^2+ \dd s^2_{S^6}) \, ,\qquad
 e^\phi\approx r^{1/4}\,;
\end{equation}
again $r=z_0-z$.

\paragraph{O8--O2.}

\begin{figure}[h]
\centering
\includegraphics[width=7cm]{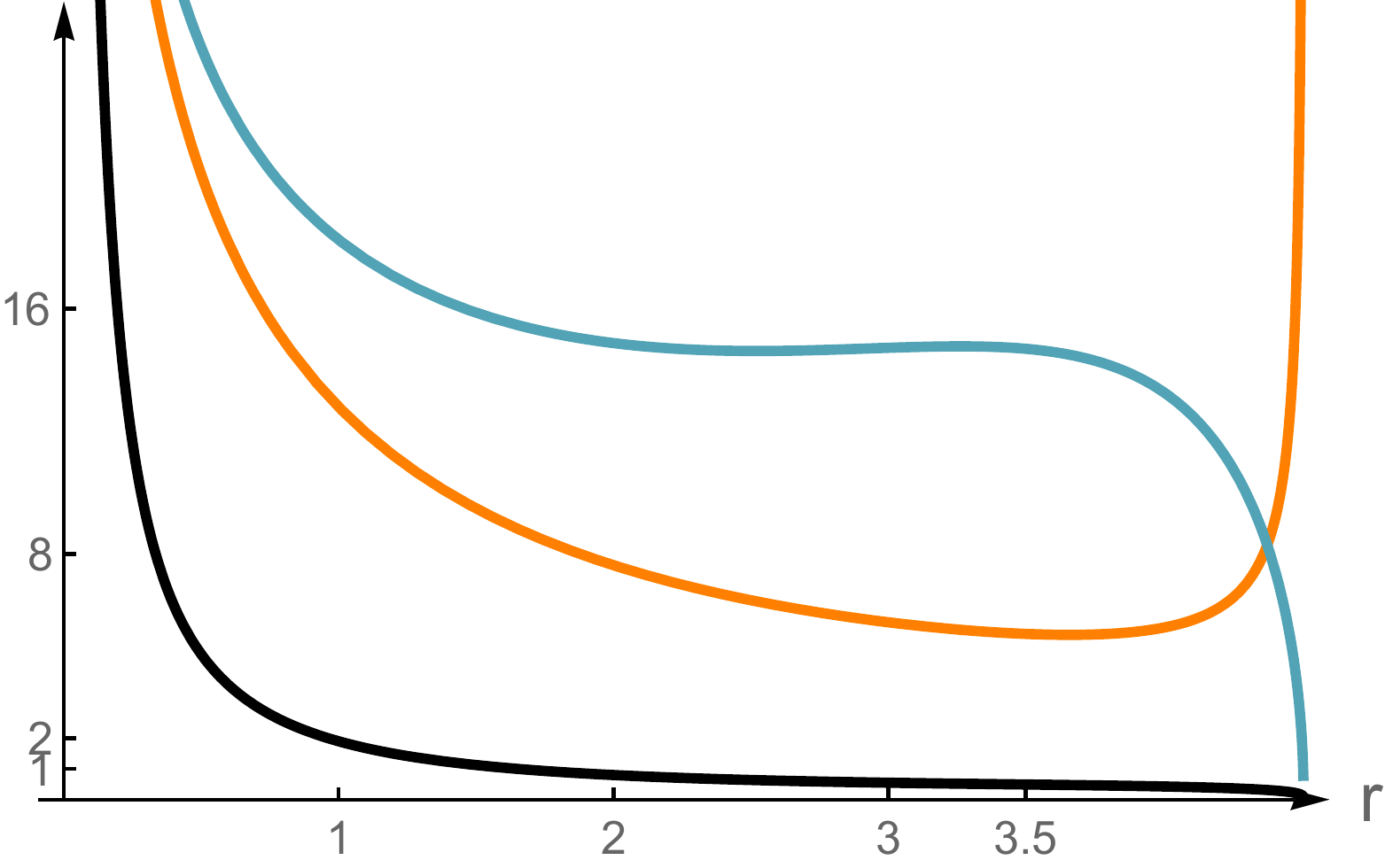}
\caption{A solution with an O8 (left extremum) and an O2 (right extremum). In orange $e^{2A}$, in blue $e^{2Q}$ and in black the dilaton $e^{\phi}$.}
\label{fig:O8-O2}
\end{figure}

Next we use the local O8 solution (\ref{eq:O8exp}) to start a numerical evolution, rather than the one from a regular point. A typical solution for $\kappa=0$ is the one in Fig.~\ref{fig:O8-O2}. The behavior on the right is the one typical of an O2. 

\begin{figure}[ht]
\centering
\includegraphics[width=7cm]{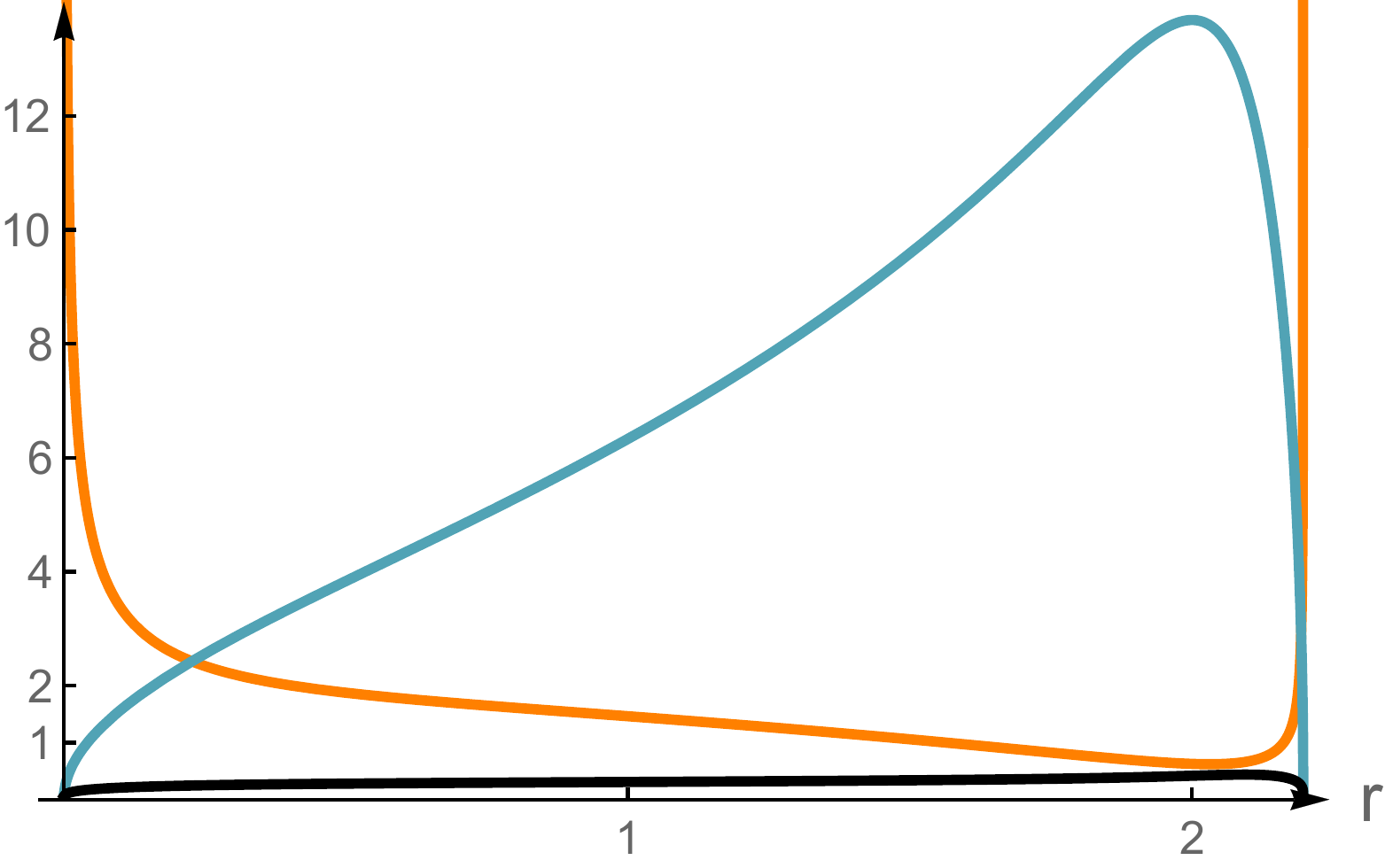}
\caption{A solution with two O2-planes. In orange $e^{2A}$, in blue $e^{2Q}$ and in black the dilaton $e^{\phi}$.}
\label{fig:O2-O2}
\end{figure}

\paragraph{O2--O2.}
Finally we use the local O2 solution (\ref{eq:O2exp}) as a starting point. We find solutions that  generically look like the one in Fig.~\ref{fig:O2-O2}. The behavior on the right is again typical of an O2, so this is a solution with two O2-planes. 

\newpage

\section*{Acknowledgements}
We would like to thank N.~Macpherson, D.~Prins, D.~Rosa for interesting discussions. N.P. and A.T. would also like to acknowledge the members of theoretical group of IPM, Tehran, for their kind hospitality while part of this work was being prepared. The work of G.D.~is supported by the Swedish Research Council (VR). The work of A.P.~is supported by the Knut and Alice Wallenberg Foundation under grant Dnr KAW 2015.0083. G.L.M. and A.T.~are supported in part by INFN. The work of N.P. was partially supported by ICTP. 

\appendix

\section{The supersymmetry system} 
\label{app:susy}

We will derive here the supersymmetry system (\ref{eq:susy}); see also footnote \ref{foot:mmpr}. We will use the formalism of \cite{10d}, which was used already in section 4.2 of that paper to rederive more quickly the Minkowski$_3$ system of \cite{haack-lust-martucci-t}

We start from decomposing the fluxes: 
\begin{equation}
	H_{\rm 10D} \equiv h e^{3A} \mathrm{vol}_3 + H \, ,\qquad  F_{\rm 10D} \equiv e^{3A} \mathrm{vol}_3 \wedge * \lambda F + F\,;
\end{equation}                                                                                                                    
they obey $d_{H_{\rm 10D}}F_{\rm 10D}=0$, which decompose as
\begin{equation}\label{eq:dec-Bianchi}
	d_H (e^{3A} * \lambda F)= - h F \, ,\qquad d_H F = 0 \,.
\end{equation}

We now look at \cite[(3.1b)]{10d}:
\begin{equation}\label{eq:LK}
    L_K g = 0 \, ,\qquad d\tilde K = \iota_K H_{\rm 10D}\,,
\end{equation}
where $K^M \equiv \overline{\epsilon_1} \gamma^M \epsilon_1 + \overline{\epsilon_2} \gamma^M \epsilon_2$, $\tilde K_M \equiv \overline{\epsilon_1} \gamma^M \epsilon_1 - \overline{\epsilon_2} \gamma^M \epsilon_2 $. From (\ref{eq:eps}), 
\begin{equation}
\tilde{K}_\mu = \frac{1}{16} e^A n_- v_\mu\,, \qquad K^\mu = \frac{1}{16} e^{-A} n_+ v^\mu\,, 
\end{equation}
where 
\begin{equation}
	n_\pm \equiv ||\eta_1||^2 \pm ||\eta_2||^2 \, ,\qquad        v_\mu \equiv \frac{1}{2} \zeta^t \gamma^0 \gamma_\mu \zeta\,;
\end{equation}
while the internal components $K^m = \tilde K_m = 0$. Notice that $v^2=0$, i.e.~$v$ is null. 

From the fact that $K$ is a Killing vector (first equation in (\ref{eq:LK})) we see that $e^{-A} n_+ \equiv c_+$ is constant. The second in (\ref{eq:LK}) reads
\begin{equation}\label{eq:dKt}
	d (e^A n_- v)= e^{-A} n_+ h \iota_v \mathrm{vol}_3 
	\,.
\end{equation}
It follows from this that $e^A n_-\equiv c_-$ is also constant. Moreover, the Killing condition (\ref{eq:Ksp}) for $\zeta$
implies 
\begin{equation}
d_3 v = - 2 \mu *_3 v\,.
\end{equation}
Since $v$ is null, one can also see $*_3 v = \iota_v \mathrm{vol}_3$. (\ref{eq:dKt}) then becomes
\begin{equation} \label{eq:c+h}
2 \mu c_- = - c_+ h \,.
\end{equation}

We now turn to \cite[(3.1a)]{10d}:
\begin{equation}\label{eq:psp10}
	d_{H_{\rm 10D}}(e^{-\phi} \Phi) = -(\tilde K\wedge + \iota_K ) F_{\rm 10D}\,.
\end{equation}
$\Phi$ is defined as $\epsilon_1 \overline{\epsilon_2}$; from (\ref{eq:eps}) we get
\begin{equation}
\Phi = - *_3 v  \wedge e^{2A} \psi_+ - v \wedge e^A \psi_- \,.
\end{equation}
Decomposing (\ref{eq:psp10}) along $v$ and $*_3v$, we then get
\begin{subequations}
\begin{align}
\label{eq:c-F} d_H(e^{A-\phi} \psi_-) &= - \frac{1}{16} c_- F\,, \\
\label{eq:c+F} d_H(e^{2A-\phi} \psi_+) - 2 \mu e^{A-\phi} \psi_- &= \frac{1}{16} c_+ e^{3A} * \lambda F\,.
\end{align}
\end{subequations} 

By hitting (\ref{eq:c+F}) with $d_H$, using (\ref{eq:dec-Bianchi}) and comparing with (\ref{eq:c-F}), we obtain $F(2\mu c_-+ c_+ h)=0$. In this paper we take $F\neq 0$, so we recover (\ref{eq:c+h}) again.

The zero-form part of (\ref{eq:c-F}) gives $c_- F_0=0$. In this paper we will focus on solutions with $F_0\neq 0$. So in this case we get $c_-=0$; in other words, the internal spinors have equal norm. In the main text we use $c_+=2$. From (\ref{eq:c+h}) we see then that $h=0$, and $H_{\rm 10D}= H$.

We are finally left with the ``pairing equations'', \cite[(3.1c,d)]{10d}. These are redundant for AdS$_d$ and Minkowski$_d$ compactifications for $d\ge 4$, but they do have a single component for Minkowski$_3$, in the case where both free indices are along spacetime \cite[Sec.~4.2]{10d}. A similar computation for AdS$_3$ gives 
\begin{equation}
	(\psi_-,F)_7 = \frac \mu2 e^{-\phi} \mathrm{vol}_7 \,.
\end{equation}
The last equation in (\ref{eq:susypair}) is a normalization, that follows easily from (\ref{eq:psi}).


\bibliography{at}
\bibliographystyle{at}

\end{document}